\documentclass[showpacs,preprintnumbers,amsmath,amssymb,floatfix]{revtex4}
\usepackage{graphicx}

\begin{document}

\title{Quantum inequalities for massless spin-3/2 field in Minkowski spacetime}
\date{\today}
\author{Bo Hu}
\email{bohu@ncu.edu.cn}\affiliation{Center for Gravity and
Relativistic Astrophysics, Department of Physics, Nanchang
University, Nanchang, 330047, China}
\author{Yi Ling}
\email{yling@ncu.edu.cn} \affiliation{Center for Gravity and
Relativistic Astrophysics, Department of Physics, Nanchang
University, Nanchang, 330047, China\\
CCAST (World Laboratory), P.O. Box 8730, Beijing,
   100080, China}
    \author{Hongbao Zhang}
    \email{hbzhang@pkuaa.edu.cn}
    \affiliation{Department of Physics, Beijing Normal University, Beijing, 100875, China\\
Department of Astronomy, Beijing Normal University, Beijing, 100875,
China}

\begin{abstract}Quantum inequalities have been established for various quantum
fields in both flat and curved spacetimes. In particular, for
spin-3/2 fields, Yu and Wu have explicitly derived quantum
inequalities for massive case. Employing the similar method
developed by Fewster and colleagues, this paper provides an explicit
formula of quantum inequalities for massless spin-3/2 field in
four-dimensional Minkowski spacetime.
\end{abstract}

\pacs{03.70.+k, 11.10.Ef}

\maketitle
\section{Introduction}
As is well known, seemingly reasonable energy conditions such as
weak energy condition play a special role in classical general
relativity. It has proved profitable to only require one or more
energy conditions satisfied by the energy momentum tensor rather
than to know the specific expression of the energy momentum tensor
for matters. For example, the singularity theorem and the positive
mass conjecture are proved under such assumptions.

However, all the pointwise energy conditions are violated in the
framework of quantum field theory. Even there exist such series of
quantum states in which the energy density at a given point may
approach arbitrary negative values. If the magnitude and duration
of such negative energy densities were unconstrained, various
exotic phenomena might occur. These result in serious
ramifications such as the violation of the second law of
thermodynamics, and the existence of traversable wormholes, and
even time machines.

Interestingly, there exist some mechanisms in quantum field theory
to restrict the extent of negative energy densities: the weighted
average of energy densities by non-negative sampling functions
satisfies the quantum weak energy inequalities, simply called
quantum inequalities. Since the pioneering work by Ford, who
obtained a quantum inequality for massless minimally-coupled scalar
field in Minkowski spacetime with a Lorentzian sampling
function\cite{Ford}, progress has been made toward generalizing to
various quantum fields in both flat and curved
spacetimes\cite{FR1,PF1,Flanagan1,PF2,FE,FT,Fewster1,Vollick,FV,
Pfenning,Flanagan2,FM,FP,YW, Fewster2,FH, Dawson}. Especially, for
spin-3/2 field, using the method developed by Fewster and
colleagues\cite{FE,FT,FV,FM}, Yu and Wu have given an explicit
derivation of quantum inequalities for massive case in
four-dimensional Minkowski spacetime\cite{YW}. As a further step
along this line, this paper will provide quantum inequalities for
massless spin-3/2 field in four-dimensional Minkowski spacetime for
arbitrary non-negative sampling functions by the same approach. The
result obtained here is
\begin{equation}
\int_{-\infty}^\infty dx^0\langle\hat{\rho}(x^0,\mathbf{x})\rangle
g^2(x^0)\geq-\frac{1}{24\pi^3}\int_0^\infty du|\tilde{g}(u)|^2u^4.
\end{equation}
Here notations and conventions follow those in \cite{SZ}.
Especially, the metric signature takes $(+,-,-,-)$, and
$\{\sigma^\mu{_{\Sigma'\Sigma}}=\frac{1}{\sqrt{2}}(I,\sigma)|\mu=0,1,2,3;\Sigma(\Sigma')=1,2\}$
with $\sigma$ Pauli matrices. In addition, the Fourier transformer
of a function $g$ is defined by
\begin{equation}
\tilde{g}(\omega)=\int_{-\infty}^\infty dx^0g(x^0)e^{-i\omega x^0}.
\end{equation}
\section{Equation of motion and energy momentum tensor for massless
spin-3/2 field from Rarita-Schwinger Lagrangian} This section will
present a brief review of the theory of massless spin-3/2 field in
four-dimensional Minkowski spacetime, which provides a concise
foundation for later work. For more details, please refer
to\cite{SZ}.

Start with Rarita-Schwinger Lagrangian\cite{SZ,RS}
\begin{equation}
\mathcal{L}=-i\sqrt{2}[\bar{\psi}^{aB'}\sigma^b{_{B'B}}\nabla_b\psi_a{^B}-\frac{1}{3}(\bar{\psi}^{aB'}\sigma_{aB'B}\nabla_b\psi^{bB}+\bar{\psi}^{aB'}\sigma_{bB'B}\nabla_a\psi^{bB})+\frac{2}{3}\bar{\psi}^{aB'}\sigma_{aB'B}\sigma^{bBC'}\sigma_{cC'C}\nabla_b\psi^{cC}],\label{Lagrangian}
\end{equation}
where the bar denotes the Hermitian conjugation. From here,
Euler-Lagrange equation leads to
\begin{equation}
\sigma^b{_{B'B}}\nabla_b\psi_a{^B}-\frac{1}{3}(\sigma_{aB'B}\nabla_b\psi^{bB}+\sigma_{bB'B}\nabla_a\psi^{bB})+\frac{2}{3}\sigma_{aB'B}\sigma^{bBC'}\sigma_{cC'C}\nabla_b\psi^{cC}=0.
\end{equation}
With the covariant derivative and the soldering form action on the
equation of motion, respectively, we have
\begin{eqnarray}
\sigma^b{_{B'B}}\nabla_b\nabla^a\psi_a{^B}&=&0, \nonumber\\
\nabla^a\psi_a{^B}&=&0,
\end{eqnarray}
where the identity
$\sigma_{aCB'}\sigma_{bD}{^{B'}}+\sigma_{bCB'}\sigma_{aD}{^{B'}}=\eta_{ab}\epsilon_{CD}$
has been employed\cite{SZ}. Taking into account Rarita-Schwinger
constraint condition, i.e.,
\begin{equation}
\sigma^a{_{B'B}}\psi_a{^B}=0,\label{RS1}
\end{equation}
the equation of motion is simplified as
\begin{equation}
\sigma^b{_{B'B}}\nabla_b\psi_a{^B}=0.\label{RS2}
\end{equation}
Eqn.(\ref{RS1}) and Eqn.(\ref{RS2}) are just our familiar
Rarita-Schwinger equations for massless spin-3/2 field\cite{SZ,RS}.
Furthermore, by Belinfante's construction and after a
straightforward calculation, the energy momentum tensor for massless
spin-3/2 field reads\cite{SZ}
\begin{equation}
T_\mathcal{B}^{ab}=-i\sqrt{2}[\frac{1}{2}(\bar{\psi}^{dD'}\sigma^{(b}{_{D'E}}\nabla^{a)}\psi_d{^E}-\nabla^{(a}\bar{\psi}^{|dD'|}\sigma^{b)}{_{D'E}}\psi_d{^E})+(\nabla_c\bar{\psi}^{(b|D'|}\sigma^{a)}{_{D'D}}\psi^{cD}-\bar{\psi}^{cD'}\sigma^{(a}{_{D'D}}\nabla_c\psi^{b)D})],\label{em}
\end{equation}
which is equivalent with that obtained by the variational
principle\cite{Zhang}, thus acts as the source of Einstein's
gravitational field equation.

It is worth noting that Rarita-Schwinger field equations are
invariant under the following gauge transformation\cite{SZ,RS}
\begin{equation}
\psi_a{^B}\rightarrow\psi_a{^B}+\nabla_a\varphi^B
\end{equation}
with
\begin{equation}
\sigma^b{_{B'B}}\nabla_b\varphi^B=0.
\end{equation}
However, the energy momentum tensor (\ref{em}) is not gauge
invariant. Thus in the following discussions we will restrict
ourselves to Coulomb gauge, i.e.,
\begin{equation}
\psi_0{^B}=0.
\end{equation}
Obviously, the energy density in Coulomb gauge is given by
\begin{equation}
\rho=T^{00}=-i\frac{\sqrt{2}}{2}(\bar{\psi}^{dD'}\sigma^0{_{D'E}}\nabla^0\psi_d{^E}-\nabla^0\bar{\psi}^{dD'}\sigma^0{_{D'E}}\psi_d{^E}).\label{density}
\end{equation}
\section{Canonical quantization and quantum inequality for massless
spin-3/2 field in Coulomb gauge} To obtain quantum inequalities for
massless spin-3/2 field in Coulomb gauge, we need first quantize
massless spin-3/2 field. A consistent massless spin-3/2 quantum
field can be constructed by the plane wave basis in Coulomb gauge
as\cite{SZ}
\begin{equation}
\hat{\psi}_a{^B}(x)=\int
d^3\mathbf{p}[a(\mathbf{p})\psi_p{_a{^B}}(x)+c^\dag(\mathbf{p})\psi_{-p}{_a{^B}}(x)],p_0>0.\label{quantum}
\end{equation}
Here the annihilation and creation operators satisfy the
anti-commutation relations as follows
\begin{eqnarray}
\{a(\mathbf{p}),a(\mathbf{p'})\}&=&0,\nonumber\\
\{a(\mathbf{p}),a^\dag(\mathbf{p'})\}&=&\delta^3(\mathbf{p}-\mathbf{p'}),\nonumber\\
\{a^\dag(\mathbf{p}),a^\dag(\mathbf{p'})\}&=&0,\nonumber\\
\{c(\mathbf{p}),c(\mathbf{p'})\}&=&0,\nonumber\\
\{c(\mathbf{p}),c^\dag(\mathbf{p'})\}&=&\delta^3(\mathbf{p}-\mathbf{p'}),\nonumber\\
\{c^\dag(\mathbf{p}),c^\dag(\mathbf{p'})\}&=&0.
\end{eqnarray}
The plane wave solutions to Rarita-Schwinger equations in Coulomb
gauge read
\begin{equation}
\psi_p{_a{^B}}(x)=\frac{1}{\sqrt{(2\pi)^3}}\frac{1}{\sqrt{2|p_0|}}\tilde{\psi}_\mu{^\Sigma}(p)(dx^\mu)_a(\varepsilon_\Sigma)^Be^{-ip_bx^b}.
\end{equation}
where
\begin{equation}
\tilde{\psi}(1,0,0,1)=(0,1,i,0)\otimes\left(\begin{array}{l l}
1\\0\end{array}\right),
\end{equation}
and
\begin{eqnarray}
\tilde{\psi}_\mu{^\Sigma}(p=e^{-\lambda},e^{-\lambda}\sin\theta\cos\varphi,e^{-\lambda}\sin\theta\sin\varphi,e^{-\lambda}\cos\theta)
&=&\tilde{\psi}_\mu{^\Sigma}(-p)\nonumber\\
&=&(\Lambda^{-1})^\nu{_\mu}L^\Sigma{_\Gamma}\tilde{\psi}_\nu{^\Gamma}(1,0,0,1)
\end{eqnarray}
with
\begin{eqnarray}
\Lambda&=&\left(\begin{array}{cccc}
         1 & 0 & 0 & 0 \\
          0 & \cos\varphi & -\sin\varphi & 0 \\
          0 & \sin\varphi & \cos\varphi & 0 \\
          0 & 0 & 0 & 1
       \end{array}\right)
\left(\begin{array}{cccc}
         1 & 0 & 0 & 0 \\
          0 & \cos\theta & 0 & \sin\theta \\
          0 & 0 & 1 & 0\\
          0 & -\sin\theta & 0 & \cos\theta
       \end{array}\right)
\left(\begin{array}{cccc}
         \cosh\lambda & 0 & 0& -\sinh\lambda \\
          0 & 1 & 0 & 0 \\
          0 & 0 & 1 & 0 \\
          -\sinh\lambda & 0 & 0 & \cosh\lambda
       \end{array}\right),\nonumber\\
L&=&\left(\begin{array}{cc}
        e^{-i\frac{\varphi}{2}} & 0 \\
          0 & e^{i\frac{\varphi}{2}}
       \end{array}\right)
\left(\begin{array}{cc}
        \cos\frac{\theta}{2} & -\sin\frac{\theta}{2} \\
          \sin\frac{\theta}{2} & \cos\frac{\theta}{2}
       \end{array}\right)
\left(\begin{array}{cc}
        e^{-\frac{\lambda}{2}} & 0 \\
          0 & e^{\frac{\lambda}{2}}
       \end{array}\right).
\end{eqnarray}

Next substituting the quantum massless spin-3/2 field
(\ref{quantum}) into Eqn.(\ref{density}) and taking the normal
order, the expectation value of the quantum energy density operator
in an arbitrary quantum state can be written as
\begin{eqnarray}
\langle\hat{\rho}\rangle&=&\langle:\hat{T}^{00}:\rangle\nonumber\\
&=&\frac{1}{2}\frac{1}{(2\pi)^3}\int
\frac{d^3\mathbf{p}}{\sqrt{2p_0}}\frac{d^3\mathbf{p'}}{\sqrt{2p'_0}}(-\sqrt{2})\nonumber\\
&&\{(p_0+p'_0)[\langle a^\dag(\mathbf{p})a(\mathbf{p'})\rangle\bar{\tilde{\psi}}^{\mu\Sigma'}(p)\sigma^0{_{\Sigma'\Sigma}}\tilde{\psi}_\mu{^\Sigma}(p')e^{i(p_b-p'_b)x^b}\nonumber\\
&&+\langle c^\dag(\mathbf{p'})c(\mathbf{p})\rangle\bar{\tilde{\psi}}^{\mu\Sigma'}(-p)\sigma^0{_{\Sigma'\Sigma}}\tilde{\psi}_\mu{^\Sigma}(-p')e^{-i(p_b-p'_b)x^b}]\nonumber\\
&&+(p_0-p'_0)[\langle
a^\dag(\mathbf{p})c^\dag(\mathbf{p'})\rangle\bar{\tilde{\psi}}^{\mu\Sigma'}(p)\sigma^0{_{\Sigma'\Sigma}}\tilde{\psi}_\mu{^\Sigma}(-p')e^{i(p_b+p'_b)x^b}\nonumber\\
&&-\langle c(\mathbf{p})a(\mathbf{p'})\rangle\bar{\tilde{\psi}}^{\mu\Sigma'}(-p)\sigma^0{_{\Sigma'\Sigma}}\tilde{\psi}_\mu{^\Sigma}(p')e^{-i(p_b+p'_b)x^b}]\}.\nonumber\\
\end{eqnarray}
Now consider the sampled energy density measured by an inertial
observer at the spatial position $\mathbf{x}$, i.e.,
\begin{eqnarray}
\langle\hat{\rho}\rangle_f&=&\int_{-\infty}^\infty
dx^0\langle\hat{\rho}(x^0,\mathbf{x})\rangle f(x^0)\nonumber\\
&=&\frac{1}{2}\frac{1}{(2\pi)^3}\int
\frac{d^3\mathbf{p}}{\sqrt{2p_0}}\frac{d^3\mathbf{p'}}{\sqrt{2p'_0}}(-\sqrt{2})\nonumber\\
&&\{(p_0+p'_0)[\langle a^\dag(\mathbf{p})a(\mathbf{p'})\rangle\bar{\tilde{\psi}}^{\mu\Sigma'}(p)\sigma^0{_{\Sigma'\Sigma}}\tilde{\psi}_\mu{^\Sigma}(p')\tilde{f}(p'_0-p_0)e^{i(\mathbf{p}-\mathbf{p'})\cdot\mathbf{x}}\nonumber\\
&&+\langle c^\dag(\mathbf{p'})c(\mathbf{p})\rangle\bar{\tilde{\psi}}^{\mu\Sigma'}(-p)\sigma^0{_{\Sigma'\Sigma}}\tilde{\psi}_\mu{^\Sigma}(-p')\tilde{f}(p_0-p'_0)e^{-i(\mathbf{p}-\mathbf{p'})\cdot\mathbf{x}}]\nonumber\\
&&+(p_0-p'_0)[\langle
a^\dag(\mathbf{p})c^\dag(\mathbf{p'})\rangle\bar{\tilde{\psi}}^{\mu\Sigma'}(p)\sigma^0{_{\Sigma'\Sigma}}\tilde{\psi}_\mu{^\Sigma}(-p')\tilde{f}(-p'_0-p_0)e^{i(\mathbf{p}+\mathbf{p'})\cdot\mathbf{x}}\nonumber\\
&&-\langle c(\mathbf{p})a(\mathbf{p'})\rangle\bar{\tilde{\psi}}^{\mu\Sigma'}(-p)\sigma^0{_{\Sigma'\Sigma}}\tilde{\psi}_\mu{^\Sigma}(p')\tilde{f}(p_0+p'_0)e^{-i(\mathbf{p}+\mathbf{p'})\cdot\mathbf{x}}]\},\nonumber\\
\end{eqnarray}
where $f$ is a non-negative sampling function. Let $f=g^2$ and
introduce a family of operators
\begin{equation}
\hat{O}_\mu{^\Sigma}(\omega)=\frac{1}{\sqrt{(2\pi)^3}}\int
\frac{d^3\mathbf{p}}{\sqrt{2p_0}}\bar{\tilde{g}}(-p_0+\omega)a(\mathbf{p})\tilde{\psi}_\mu{^\Sigma}(p)e^{-i\mathbf{p}\cdot\mathbf{x}}+\bar{\tilde{g}}(p_0+\omega)c^\dag(\mathbf{p})\tilde{\psi}_\mu{^\Sigma}(-p)e^{i\mathbf{p}\cdot\mathbf{x}}.
\end{equation}
Using
\begin{equation}
\sqrt{2}\bar{\tilde{\psi}}^{\mu\Sigma'}(-p)\sigma^0{_{\Sigma'\Sigma}}\tilde{\psi}_\mu{^\Sigma}(-p)=\sqrt{2}\Lambda^0{_\nu}\bar{\tilde{\psi}}^{\mu\Sigma'}(1,0,0,1)\sigma^\nu{_{\Sigma'\Sigma}}\tilde{\psi}_\mu{^\Sigma}(1,0,0,1)=2p_0.
\end{equation}
and $\tilde{g}(-\omega)=\bar{\tilde{g}}(\omega)$, it can be shown
that
\begin{eqnarray}
-\sqrt{2}\bar{\hat{O}}^{\mu\Sigma'}(\omega)\sigma^0{_{\Sigma'\Sigma}}\hat{O}_\mu{^\Sigma}(\omega)&=&Z(\omega)+\frac{1}{(2\pi)^3}\int
\frac{d^3\mathbf{p}}{\sqrt{2p_0}}\frac{d^3\mathbf{p'}}{\sqrt{2p'_0}}(-\sqrt{2})\nonumber\\
&&\{\tilde{g}(-p_0+\omega)\bar{\tilde{g}}(-p'_0+\omega)a^\dag(\mathbf{p})a(\mathbf{p'})\bar{\tilde{\psi}}^{\mu\Sigma'}(p)\sigma^0{_{\Sigma'\Sigma}}\tilde{\psi}_\mu{^\Sigma}(p')e^{i(\mathbf{p}-\mathbf{p'})\cdot\mathbf{x}}\nonumber\\
&&-\tilde{g}(p_0+\omega)\bar{\tilde{g}}(p'_0+\omega)c^\dag(\mathbf{p'})c(\mathbf{p})\bar{\tilde{\psi}}^{\mu\Sigma'}(-p)\sigma^0{_{\Sigma'\Sigma}}\tilde{\psi}_\mu{^\Sigma}(-p')e^{-i(\mathbf{p}-\mathbf{p'})\cdot\mathbf{x}}\nonumber\\
&&+\tilde{g}(-p_0+\omega)\bar{\tilde{g}}(p'_0+\omega)a^\dag(\mathbf{p})c^\dag(\mathbf{p'})\bar{\tilde{\psi}}^{\mu\Sigma'}(p)\sigma^0{_{\Sigma'\Sigma}}\tilde{\psi}_\mu{^\Sigma}(-p')e^{i(\mathbf{p}+\mathbf{p'})\cdot\mathbf{x}}\nonumber\\
&&+\tilde{g}(p_0+\omega)\bar{\tilde{g}}(-p'_0+\omega)c(\mathbf{p})a(\mathbf{p'})\bar{\tilde{\psi}}^{\mu\Sigma'}(-p)\sigma^0{_{\Sigma'\Sigma}}\tilde{\psi}_\mu{^\Sigma}(p')e^{-i(\mathbf{p}+\mathbf{p'})\cdot\mathbf{x}}\},\nonumber\\
\end{eqnarray}
and
\begin{equation}
-\sqrt{2}\sigma^0{_{\Sigma'\Sigma}}[\bar{\hat{O}}^{\mu\Sigma'}(\omega)\hat{O}_\mu{^\Sigma}(\omega)+\hat{O}_\mu{^\Sigma}(\omega)\bar{\hat{O}}^{\mu\Sigma'}(\omega)]=Z(-\omega)+Z(\omega)
\end{equation}
 with
\begin{equation}
Z(\omega)=\frac{1}{(2\pi)^3}\int
d^3\mathbf{p}|\tilde{g}(p_0+\omega)|^2.
\end{equation}
Employing the identity\cite{FV,FM}
\begin{equation}
(p_0+p'_0)\tilde{f}(p_0-p'_0)=\frac{1}{\pi}\int_{-\infty}^\infty
d\omega\tilde{g}(p_0-\omega)\bar{\tilde{g}}(p'_0-\omega)\omega,
\end{equation}
we obtain
\begin{eqnarray}
\langle\hat{\rho}\rangle_f&=&\frac{1}{2\pi}\int_{-\infty}^\infty
d\omega[\langle-\sqrt{2}\bar{\hat{O}}^{\mu\Sigma'}(\omega)\sigma^0{_{\Sigma'\Sigma}}\hat{O}_\mu{^\Sigma}(\omega)\rangle-Z(\omega)]\omega\nonumber\\
&=&\frac{1}{2\pi}\{\int_{-\infty}^0d\omega[Z(-\omega)+Z(\omega)
+\langle\sqrt{2}\hat{O}_\mu{^\Sigma}(\omega)\sigma^0{_{\Sigma'\Sigma}}\bar{\hat{O}}^{\mu\Sigma'}(\omega)\rangle-Z(\omega)]\omega\nonumber\\
&&+\int_0^\infty
d\omega[\langle-\sqrt{2}\bar{\hat{O}}^{\mu\Sigma'}(\omega)\sigma^0{_{\Sigma'\Sigma}}\hat{O}_\mu{^\Sigma}(\omega)\rangle-Z(\omega)]\omega\}\nonumber\\
&=&\frac{1}{2\pi}\{\int_{-\infty}^0d\omega[Z(-\omega)
+\langle\sqrt{2}\hat{O}_\mu{^\Sigma}(\omega)\sigma^0{_{\Sigma'\Sigma}}\bar{\hat{O}}^{\mu\Sigma'}(\omega)\rangle]\omega\nonumber\\
&&+\int_0^\infty
d\omega[\langle-\sqrt{2}\bar{\hat{O}}^{\mu\Sigma'}(\omega)\sigma^0{_{\Sigma'\Sigma}}\hat{O}_\mu{^\Sigma}(\omega)\rangle-Z(\omega)]\omega\}.
\end{eqnarray}
By \{$\hat{O}_0{^\Sigma}(\omega)=0|\Sigma=1,2\}$ in Coulomb gauge,
we have
\begin{eqnarray}
\langle\sqrt{2}\hat{O}_\mu{^\Sigma}(\omega)\sigma^0{_{\Sigma'\Sigma}}\bar{\hat{O}}^{\mu\Sigma'}(\omega)\rangle&=&-[\langle\hat{O}_1{^1}(\omega)\bar{\hat{O}}_1{^1}(\omega)\rangle+\langle\hat{O}_1{^2}(\omega)\bar{\hat{O}}_1{^2}(\omega)\rangle\nonumber\\
&&+\langle\hat{O}_2{^1}(\omega)\bar{\hat{O}}_2{^1}(\omega)\rangle+\langle\hat{O}_2{^2}(\omega)\bar{\hat{O}}_2{^2}(\omega)\rangle\nonumber\\
&&+\langle\hat{O}_3{^1}(\omega)\bar{\hat{O}}_3{^1}(\omega)\rangle+\langle\hat{O}_3{^2}(\omega)\bar{\hat{O}}_3{^2}(\omega)\rangle]\leq0,\nonumber\\
\end{eqnarray}
and similarly
\begin{equation}
\langle-\sqrt{2}\bar{\hat{O}}^{\mu\Sigma'}(\omega)\sigma^0{_{\Sigma'\Sigma}}\hat{O}_\mu{^\Sigma}(\omega)\rangle\geq0
\end{equation}
hold for arbitrary quantum states. Therefore
\begin{eqnarray}
\langle\hat{\rho}\rangle_f&\geq&\frac{1}{2\pi}\{\int_{-\infty}^0d\omega
Z(-\omega)\omega-\int_0^\infty d\omega Z(\omega)\omega\}\nonumber\\
&=&-\frac{1}{\pi}\int_0^\infty d\omega Z(\omega)\omega\nonumber\\
&=&-\frac{1}{2\pi^3}\int_0^\infty d\omega\omega\int_0^\infty
dp_0p_0^2|\tilde{g}(p_0+\omega)|^2\nonumber\\
&=&-\frac{1}{2\pi^3}\int_0^\infty dp_0\int_{p_0}^\infty
du(u-p_0)p_0^2|\tilde{g}(u)|^2\nonumber\\
&=&-\frac{1}{2\pi^3}\int_0^\infty
du\int_0^udp_0(u-p_0)p_0^2|\tilde{g}(u)|^2\nonumber\\
&=&-\frac{1}{24\pi^3}\int_0^\infty du|\tilde{g}(u)|^2u^4.
\end{eqnarray}
\section{Discussions}
In summary, based on the technique developed by Fewster and
colleagues\cite{FE,FT,FV,FM}, we have obtained an explicit formula
of quantum inequalities for massless spin-3/2 field by arbitrary
non-negative sampling function in Coulomb gauge. It is worth noting
that the bound here is weaker, by a factor of $4$, than that
obtained by taking massless limit of quantum inequalities for
massive spin-3/2 field in four-dimensional Minkowski
spacetime\cite{YW}. This seems to originate from the fact that massive
spin-3/2 field has the degrees of freedom with four times as many as
massless one. Thus like bosonic fields, the quantum inequalities
derived so far for fermionic fields in four-dimensional Minkowski
spacetime can also be written in terms of a unified form\cite{FM,YW}
\begin{equation}
\int_{-\infty}^\infty dx^0\langle\hat{\rho}(x^0,\mathbf{x})\rangle
g^2(x^0)\geq-\frac{s}{48\pi^3}\int_m^\infty
du|\tilde{g}(u)|^2u^4Q_3^F(\frac{u}{m}),
\end{equation}
where $s$ denotes the degrees of freedom for fields, and
\begin{equation}
Q_3^F(z)=4(1-\frac{1}{z^2})^\frac{3}{2}-3[(1-\frac{1}{z^2})^\frac{1}{2}(1-\frac{1}{2z^2})-\frac{1}{2z^4}\ln(z+\sqrt{z^2-1})],
\end{equation}
which is replaced by $\lim_{z\rightarrow\infty}Q_3^F(z)=1$ in the
massless case.

We conclude with an important question. Different from
electromagnetic field, there is no gauge invariant energy momentum
tensor for massless spin-3/2 field, which is also shared by linear
gravitational field indeed. Since we here choose a particular gauge,
it is a natural question whether the quantum inequalities obtained
here for massless spin-3/2 field is gauge independent. However, the
answer is not obvious, thus worthy of further investigation, which
is expected to be reported elsewhere.
\section*{Acknowledgements}
We are particularly grateful to Prof. H. Yu for his helpful
communication. H. Zhang would like to thank the Center for Gravity
and Relativistic Astrophysics at Nanchang University for its
hospitality during his present visit. B. Hu's work is partly
supported by NSFC(No.10505011), Y. Ling's work is partly supported
by NSFC (Nos.10405027 and 10205002) and SRF for ROCS, SEM, and H.
Zhang's work is supported in part by NSFC(Nos.10373003 and
10205002).


\begin{thebibliography}{99}
\bibitem{Ford}L. H. Ford, Phys. Rev. D 43: 3972(1991).
\bibitem{FR1}L. H. Ford and T. A. Roman, Phys. Rev. D 55: 2082(1997).
\bibitem{PF1}M. J. Pfenning and L. H. Ford, Phys. Rev. D 55:
4813(1997).
\bibitem{Flanagan1}E. E. Flanagan, Phys. Rev. D 56: 4922(1997).
\bibitem{PF2}M. J. Pfenning and L. H. Ford, Phys. Rev. D 57:
3489(1998).
\bibitem{FE}C. J. Fewster and S. P. Eveson, Phys. Rev. D 58:
084010(1998).
\bibitem{FT}C. J. Fewster and E. Teo, Phys. Rev. D 59: 104016(1999).
\bibitem{Fewster1}C. J. Fewster, Class. Quant. Grav. 17: 1897(2000).
\bibitem{Vollick}D. N. Vollick,  Phys. Rev. D 61: 084022(2000).
\bibitem{FV}C. J. Fewster and R. Verch, Commun. Math. Phys. 225:
331(2002).
\bibitem{Pfenning}M. J. Pfenning, Phys. Rev. D 65: 024009(2002).
\bibitem{Flanagan2}E. E. Flanagan, Phys. Rev. D 66: 104007(2002).
\bibitem{FM}C. J. Fewster and B. Mistry, Phys. Rev. D 68:
105010(2003).
\bibitem{FP}C. J. Fewster and M. J. Pfenning, J. Math. Phys.
44: 4480(2003).
\bibitem{YW}H. Yu and P. Wu, Phys. Rev. D 69: 064008(2004).
\bibitem{Fewster2}C. J. Fewster, Phys. Rev. D 70: 127501(2004).
\bibitem{FH}C. J. Fewster and S. Hollands, Rev. Math. Phys. 17:
577(2005).
\bibitem{Dawson}S. P. Dawson, Class. Quantum Grav. 23: 287(2006).
\bibitem{SZ}F. Sun and H. Zhang, hep-th/0601011.
\bibitem{RS}W. Rarita and J. Schwinger, Phys. Rev. 60: 61(1941).
\bibitem{Zhang}H. Zhang, Commun. Theor. Phys. 44: 1007(2005).
\end{thebibliography}
\end{document}